\documentclass[runningheads]{llncs}
\usepackage{graphicx}
\begin{document}
\title{Toward Building Safer Smart Homes for the People with Disabilities}
\author{Shahinur Alam \inst{1}
\orcidID{0000-0001-9178-7328} \and
Md Sultan Mahmud \inst{1} \and
Mohammed Yeasin \inst{1}}

\authorrunning{Shahinur. et al.}

\institute{The University of Memphis, TN, USA \\
\email{\{salam,mmahmud,myeasin\}@memphis.edu}}

\maketitle              % typeset the header of the contribution
\begin{abstract}
Situational awareness is a critical foundation for the protection of human life/properties and is challenging to maintain for people with disabilities (i.e., visual impairments and limited mobility). In this paper, we present a dialog enabled end-to-end assistive solution called “SafeAccess” to build a safer smart home by providing situational awareness. The key functions of SafeAccess are: - 1) monitoring homes and identifying incoming persons; 2) helping users in assessing incoming threats (e.g., burglary, robbery, gun violence); and, 3) allowing users to grant safe access to homes for friends/families. In this work, we focus on building a robust model for detecting and recognizing person, generating image descriptions, and designing a prototype for the smart door. To interact with the system, we implemented a dialog enabled smartphone app, especially for creating a personalized profile from face images or videos of friends/families. A Raspberry pi connected to the home monitoring cameras captures the video frames and performs change detection to identify frames with activities. Then, we detect human presence using Faster r-cnn and extract faces using Multi-task Cascaded Convolutional Networks (MTCNN). Subsequently, we match the detected faces using FaceNet/support vector machine (SVM) classifiers. The system notifies users with an MMS containing the name of incoming persons or as “unknown,” scene image, facial description, and contextual information. The users can grant access or call emergency services using the SafeAccess app based on the received notification. Our system identifies persons with an F-score 0.97 and recognizes items to generate image description with an average F-score 0.97.

\end{abstract}

\section{Introduction}
According to Uniform Crime Reporting (UCR) Program conducted by FBI\footnote{https://ucr.fbi.gov/crime-in-the-u.s/2018/crime-in-the-u.s.-2018}, an estimated 1,206,836 violent crimes (murder, nonnegligent manslaughter, rape, robbery, and aggravated assault) and 10,208,334 property crimes (e.g., burglary, larceny-theft, motor vehicle theft, and arson) occurred nationwide in 2018. FBI report showed our homes are unsafe even though we think it is comforting and secure. The homes are even more unsafe for people with disabilities, such as visually impaired, paralyzed/ partially paralyzed, deaf, etc. The National Crime Victimization Survey (NCVS) reported that the rate of violent crime against people with disabilities was more than three times the rate for persons without disabilities from 2009 to 2014. It is very challenging to prevent these instantaneous crimes (e.g., burglary, robbery, theft, gun violence) that happen against persons/homes. However, if an intelligent monitoring system can identify suspicious persons and helps user/emergency services in assessing the incoming threats, they can take necessary action quickly that will help to reduce the severity of the victimization and will make homes safer. 

With the recent advancement in technologies, researchers and companies have developed home security solutions (described in Problem Background section) that can detect security breaches. However, these systems cannot identify who is entering the premises, especially whether it is friends/families/caregivers or burglars/robbers/unknown person. Hence, those systems could not distinguish incoming threats vs. normal activities. In addition, the users need to monitor the scene from their smartphone to understand it, which is impractical for the people with disabilities. To understand how much safe and secured people with/without disabilities feel when they are at home alone, we conducted a survey \footnote{https://github.com/salammemphis/Documents/blob/master/survey\%20questionnaire.docx} in Amazon mechanical Turk \footnote{https://www.mturk.com/}  and followed participatory design approaches (See Participatory Design). The survey results (see figure \ref{survey}), crime reports, the economic burden for the disabilities, and property losses motivated us and confirmed the need of developing an intelligent assistive solution to enhance safety, security, comfort, and quality of life of the people with or without disabilities. Although there are numerous commercial solutions available to increase home security, those solutions have some major shortcomings (discussed in Problem Background section). Hence, we propose a robust and intelligent assistive solution to build safer smart homes. The novelty of this work are: 1) building computationally efficient and energy-aware situational awareness system to monitor houses, 2) developing a model for identifying persons in front of the door and around the house, 3) building a convolutional neural network (CNN) based model for generating image description to assess incoming threats, 4) designing a user interface for interacting with the system with a minimal cognitive effort, 5) implementing a discreet and intuitive feedback mechanism, 6) designing a prototype for a cost-effective smart door with a remote control  mechanism.

The houses are equipped with cameras placed in strategic locations such as front door, back door, driveway, basement, garage, etc. A personalized profile (see Personal Profile section) is created for each home using SafeAccess app that contains demographic information and pictures of friends/family members. SafeAccess monitors the houses 24/7 and identifies incoming persons by matching face images with the personal profile. Besides, it categorizes persons into groups such as friends/families/intruders and generates an image description from contextual information and facial descriptions. A concise message is generated from the image description to asses incoming threats. The sample descriptions are like:- 1) “A friend, named John with black hair, beard, and mustache wearing eyeglasses in front of the entrance”; 2) “An unknown person who has a gun and wearing a mask at the back door”. The system notifies users via Multimedia Messaging Service (MMS) and phone calls. The users can grant access to home using the SafeAccess app.

\section{Problem Background}
Building a safer smart home has been a topic of active research for decades and received an upsurge of interest recently. Although, people install smart and safety equipment such as smoke detector, poison detector, glass-break sensors, water leakage detector, central control unit for electric kits, smart doors, etc. and turn a normal home to smart one. However, an essential safety component that is missing in current smart homes is an intelligent integrated solution- to monitor who (friends/families vs burglars/robbers/unknown) is entering homes and to enable people with disabilities to grant access safely and remotely for friends and families. Identifying persons and assessing incoming threats for homes is challenging for the people with or without disability since it requires continuous monitoring by a human observer. Sulman and colleagues \cite{sulman2008effective}) found in a study that when the number of monitoring displays increases, human performance deteriorates. They reported that a human observer missed 20\% of the event while monitoring four surveillance display. However, when they increased the number of the display window to nine, missing rates rose to 60\%.

	In last decade, numerous application has been developed to assist people with disabilities in navigation  \cite {gude2013blind}, expression detection \cite{anam2014expression} , currency recognition  \cite {looktel}, ambient awareness  \cite {ahmed2018image}, object recognition \cite{alam2015map,kao1996object,mapelli1997role,chincha2011finding,bigham2010vizwiz}. However, developing an automated system to identify persons and to assess incoming threats to homes did not receive considerable attention from the researchers. Although, the recent advancement in Machine Learning and Computer Vision, especially Convolutional Neural Network (CNN) has made the object detection \cite{krizhevsky2012imagenet,girshick2014rich,girshick2015fast,sermanet2013overfeat,he2016deep} person recognition \cite {sun2015deepid3,nezami2018face} and image captioning \cite {yagcioglu2015distributed,xu2015show,ushiku2015common,vedantam2015cider,verma2014im2text,vinyals2015show}     task robust and efficient compared to the last decade. Nevertheless, those technologies have not been used widely to build assistive solutions, especially for rising situational awarness. Moreover, the available automated commercial security solutions such as ADT  \cite {ADT}, Vivint  \cite {Vivint}, SimpliSafe  \cite {Simplisafe} , Frontpoint  \cite {FrontPointSecurity}, Honeywell  \cite {Honeywell}, etc. depends only on motion sensors to detect activities and security breaches. For example, when someone enters in the monitoring zone, it sends a push notification to the users. Then the users need to see the scene image to find out who (friends/families/caregivers or intruders) is there. It is not suitable for people with disabilities, especially for people with vision impairments. These systems do not provide any intelligent feedbacks to assess incoming threats and do not monitor who is entering and leaving home. In, addition, those systems are not designed to provide seamless access to homes to friends/families. 
\section{Participatory Design}
\begin{figure}
\includegraphics[width=\textwidth]{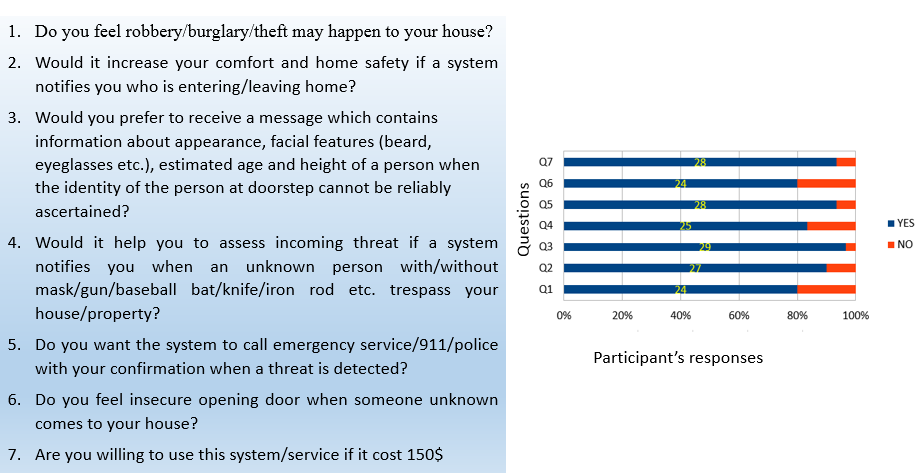}
\caption{The participants were asked seven questions related to home safety and security, and they responded with ``YES/NO" options. The “blue color” and “yellow color” represents percentage of participants agreed and disagreed with the statements respectively. Since majority of the participants agreed with the statements, the SafeAccess system will increase the safety and comfort of the people with disabilities.} 
\label{survey}
\end{figure}

The primary objective of a ``Participatory Design" is collecting the system's functional requirements and discovering issues related to accessibility and usability. The participatory design emphasizes on ``design for the users" and ``design with the users." Before designing an assistive solution, one must answer the following questions: 1) What are the offered services? 2) Who are the targeted users? 3)What type of disability they have and the severity? 4) How are their technical adaptability and cognitive ability? 5) What would be the mode for system interaction and feedback? 6) How much system costs? 7) How much they can afford? To answer those questions, we have conducted a survey  with a set of 15 questionnaires \footnote{https://github.com/salammemphis/Documents/blob/master/survey\%20questionnaire.docx}. Thirty people (26 males, 4 female; 5 partially paralyzed,8 visually impaired, 6 has hearing disability and 11 has other disabilities) participated in that survey. We asked seven questions to understand the necessity and impact of our work to enhance the safety of a house. The participant responded with binary (e.g., YES, NO) options and the survey results showed (see figure \ref{survey}) that our system will increase the safety and comfort of people with disabilities by raising situational awareness.  The rest of the questions were designed to collect functional requirements. The collected requirements from participatory designs are:
\begin{enumerate}
	\item Monitoring smart homes and identifying incoming persons
	\item Sending description about the incoming person and what they are carrying (image description) to assess threats
	\item Allowing users to speak with the person who is at the entrance
	\item Allowing users to grant access to homes for friends and families
	\item Providing the semi-automated option to call emergency services/911
\end{enumerate}

\begin{figure}
\includegraphics[width=\textwidth]{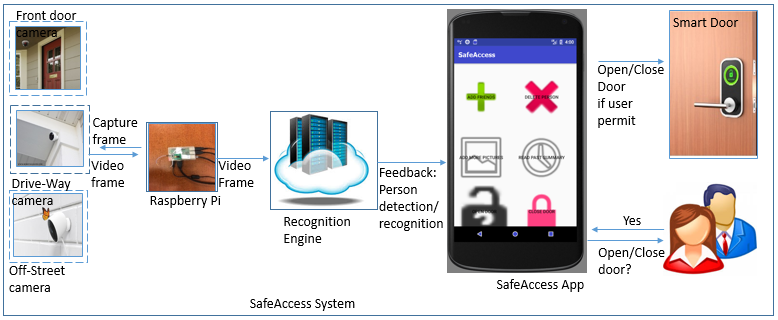}
\caption{Functional flow diagram of SafeAccess: A raspberry pi connected to the cameras captures and encrypts the real-time streams and sends them to the Image Description Generation Unit. Image Description generator detects and identifies incoming persons from the received frame and outputs an image description. The users can decide to grant access to home based on receive feedback using SafeAccess app} 
\label{safeacce_arc}
\end{figure}
\section{System Design}
Design and implementation of technology solutions require an understanding of user's needs as set by their disability and their ability to perform the system-aided task with minimal cognitive effort. Most of the smartphone-based assistive solutions suffer from severe criticism related to accessibility and usability \cite{dawe2006desperately}). For example, a touchscreen-based app is useful for sighted people, but the visually impaired individual has difficulty to use it. On the other hand, an app with a voiceover interface is convenient for visually impaired, but people with hearing disabilities cannot use it. Therefore, we have designed the user interface with two types of interaction modes. We followed iOS human interface guidelines \cite{iOS2019}  and Apple Accessibility Programming Guide \cite{iOSaccessibility} to make SafeAccess accessible for both sighted and visually impaired users. To keep the design simple and make the system effective, we used participatory design to acquire the functional requirements, design thinking to decide the aesthetic as well as usability, and system thinking to optimize the implementation with a proper feedback design. Considering the financial affordability of the users, we have developed SafeAccess with two modes: - 1)   standalone mode 2) integrated mode. In standalone mode, the computational unit is a low-cost raspberry pi (price \$35) and the system offers very limited features (person identification and granting access). We have developed a CPU and Memory efficient person recognition model (see Person Recognition) for this mode. Standalone mode is designed for people with limited family income. On the other hand, the integrated mode has an expensive server with NVIDIA RTX 2080 TI GPU and offers full features (person identification, image description generation and granting access). SafeAccess has been compartmentalized into four key components: -1) Image Description Generation module 2) Personal profile creation module 3) Prototype design for smart door 4) Feedback module. The system architecture is shown in figure \ref{safeacce_arc}.  
 
\subsection{Image Description Generation Module}
\begin{figure}
\includegraphics[width=\textwidth]{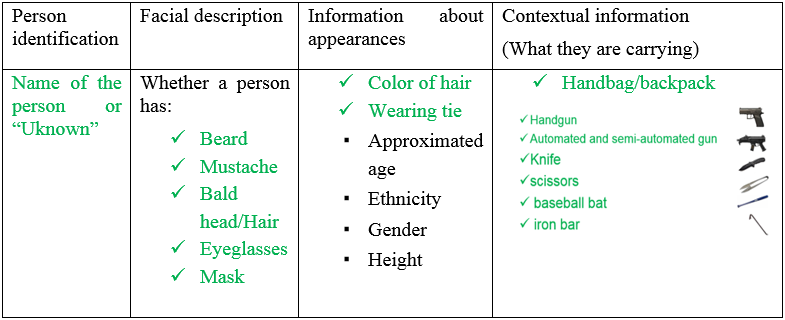}
\caption{Information included in the image description} 
\label{pd_item}
\end{figure}

\begin{figure}
\includegraphics[width=\textwidth]{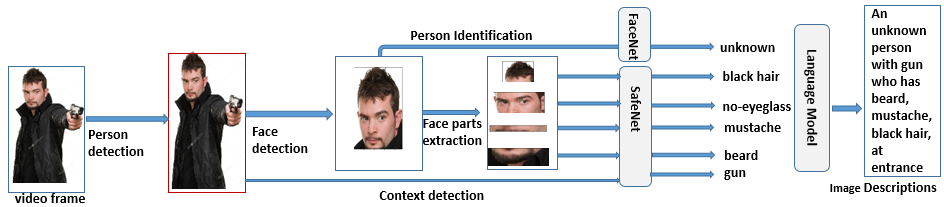}
\caption{Workflow for generating image descriptions} 
\label{wf_img}
\end{figure}

The image description is generated to present a visual summary of a scene to the visually impaired individuals so that they can understand the scene and assess incoming threats. Before generating an image description, first, we collected a list of information that helps to assess incoming threats and identify a person reliably. According to the UC Berkeley Police Department \footnote{https://ucpd.berkeley.edu/campus-safety/report-crime/describe-suspect}, survey results and outcomes from participatory design, an image description to identify a suspicious person may have information about:-1) Name of incoming person or “Unknown” 2) Facial description of incoming persons such as whether a person has a beard, mustache, etc. 3) Information about appearance 4) Contextual information (What they are carrying). The detailed list collected from participatory design is shown in figure \ref{pd_item}. The items marked with``green" color have been included in the image descriptions and others will be included in a future version of SafeAccess. 
The workflow of generating image descriptions is shown in figure \ref{wf_img}. The first step in this process is finding the presence of a human in the monitoring zone. Then the face detection, extraction, and recognition are performed. Second, face parts are extracted (see Data Collection and algorithm 2) from detected faces and a facial description is generated by classifying individual face parts. We have developed a convolutional neural network-based model called “SafeNet” (see Model Development) to extract information about facial properties, appearances, and context (what items have with them). The color of the hair has been categorized into three groups (black, brown, white) and determined by calculating the intensity histogram from the head area. The location of the detected person is obtained based on the source camera of the video frames. Finally, a semantically and syntactically meaningful image description is generated from class labels/words obtained from recognition outcomes. Synthesizing a concise and naturally flexible sentence from words generated from a complex scene (sample shown in figure \ref{pd_out}.a) with multiple persons/items is very challenging. In order to solve this problem, we are evaluating two language models LSTM\cite{hochreiter1997long} and BERT \cite{devlin2018bert}. The detailed descriptions of each component in the workflow are presented in the following sections.
\subsubsection{Person Detection:} The first task in the workflow (shown in figure \ref{wf_img}) is finding the presence of humans from camera streams because the state-of-the-art face recognition algorithms fail to recognize a person if the front view of that person is not visible to the cameras. In those extreme scenarios, at least, we need to notify users that someone is about to enter the premises. We explored state-of-the-art methods and selected Faster r-cnn \cite{ren2015faster}  as a person detection model because of its promising accuracy (see Results and Discussion section) and speed. Faster r-cnn with ResNet50 \cite{he2016deep}  has been trained on the dataset PASCALVOC12 \cite{everingham2015pascal} for 300 epochs. It took 74 hours with NVIDIA GPU GTX 1080. 
\subsubsection{Person Identification:} The next step in the pipeline (shown figure \ref{wf_img}) towards generating image description is identifying incoming persons and grouping them into friends/families/caregiver vs. unknown. We have explored both classical machine learning and deep learning methods to detect faces and identify a person. Classical methods require low computational resources and can run on low form-factor devices like raspberry pi and suitable for the Standalone mode. On the other hand, the deep learning-based model requires computationally heavy and expensive resources like GPU and applicable for the Integrated mode. The face detection and extraction for the standalone and integrated mode have been performed using Viola-jones \cite{viola2001rapid} and MTCNN \cite{zhang2016joint} respectively.
\begin{figure}
\includegraphics[width=\textwidth]{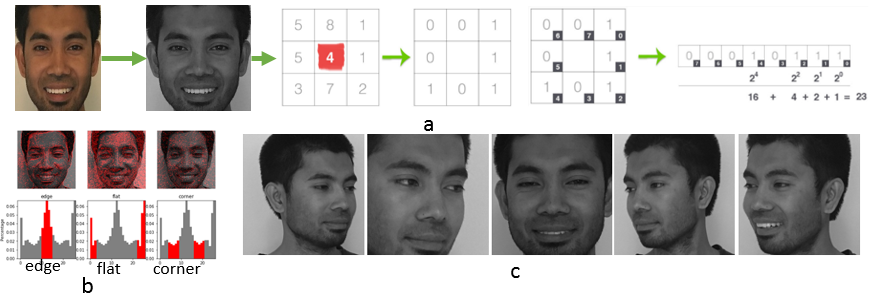}
\caption{ a) process flow to extract LBP features from images  b) sample pictures with local features such as edge, corner, etc. c) Pictures of a person from a different view included in the profile to make model robust against changes in view angle and position} 
\label{lbp_fe}
\end{figure}
	
	In order to develop a person recognition model for the standalone mode, we have explored and evaluated \textbf{(see table 4)} methods such as Eigenface \cite{turk1991face} , FisherFace \cite{belhumeur1997eigenfaces}. EigenFace is a Principal Component Analysis based model that projects the higher dimensional data into lower-dimensional space. The drawbacks of these methods are that it finds the directions with the highest variance and very error-prone to noise. Some discriminative information may be lost because it does not consider classes. Fisherface is a Linear Discriminative Analysis (LDA) based method that performs class-specific dimension reductions. Fisherface works fine in the controlled environment and with a large training sample. However, real-world scenarios are not perfect and have a limited option to control geometric and photometric information. Hence, we need to develop a model that works robustly in natural settings and as well as with a minimal number of training samples. Ojala and his colleagues \cite{ojala2002multiresolution} presented a scale and rotation invariant texture classification method using a circular local binary pattern (LBP), which requires low computational resources. LBP extracts features such as textures and shapes from the small local region of an image and has a low dimension inherently. Local features detected by the LBP include spots, flat areas, edges, edge ends, curves (Shown in figure \ref{lbp_fe}.b). The local features are detected based on the number of neighbors that have higher or lower values than the central pixel. For example, if all neighboring pixels are higher/lower than central pixels, then that part of the image is flat. Considering the computational cost and robustness against the changes in intensities, scale, and rotation, we used LBP features for classification. First, we extracted LBP features from the face images with a radius 1 and sampling point 8. Then we calculated histograms from LBP features and trained support vector machine (SVM) classifier to identify a person. The process flow are shown in figure \ref{lbp_fe}.a and involved steps are:-1) Convert RGB image to grayscale. 2) Choose a radius and number of sampling points for the circular neighborhood. 3) Slide the window over the image and calculate binary patterns/codes for each pixel. 4) Convert uniform and non-uniform patterns into a decimal number and create histograms. 5) Train SVM with calculated LBP histograms.

\subsection{Personal Profile Creation Module}
\begin{figure}
\includegraphics[width=\textwidth]{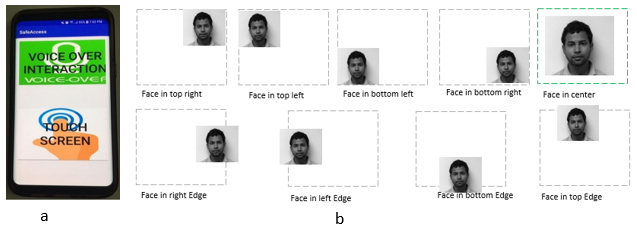}
\caption{ a) SafeAccess app with voice over and touch screen interaction mode b) Guidance on face positions in the camera window during profile creation to make sure captured faces are not cropped} 
\label{faceposition}
\end{figure}

\begin{figure}
\includegraphics[width=\textwidth]{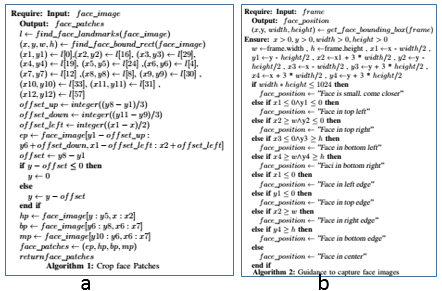}
\caption{a) Algorithm to crop face parts b) Algorithm to generate feedback for guiding users during profile creation to make sure faces are not cropped } 
\label{algs}
\end{figure}

 To identify a person, first, the recognition model needs to be trained with face images of friends/families. The “Personal Profile” is a repository which contains demographic information (Name, Email, phone number) and face images of friends/families. It is advised to include face images with various expressions (Joy, Sad, Surprise, Fear, Contempt, Disgust), orientation, and poses (see figure \ref{lbp_fe}.c) so that the system can recognize a person robustly from different view angle, position, and distances. Since smartphones are widely available, we developed a smartphone app, ``SafeAccess" to enlist a person to the profile. The app allows users to collect face images from the video clip/camera preview. To capture face images from different view systems guides the user to rotate smartphones around the face from left to right or right to left. The key challenges are: 1) enabling people with disabilities, especially visually impaired individuals to take pictures of themselves and their friends and families. 2) Providing guidance such as information about the position and size of a face in the camera window to make sure faces are not cropped, 3) Providing information about rotational speed to prevent blur in the picture. To address these challenges, first, we added both voiceover and touch screen interaction (see figure \ref{faceposition}.a) in the SafeAccess app so that users can choose an interaction mode based on their disabilities. 
 Second, a face detector \cite{viola2001rapid} has been incorporated to make the image acquisition convenient. It will make sure the selected view has faces in the right position by providing feedback such as ``Face in top right", “Face in center” etc. (sample is shown in figure \ref{faceposition}.b).  Moreover, if the person is far away from the camera and the size of the detected face is very small, then the system guides them to come closer. The underlying logic and algorithm is shown in figure \ref{algs}.b where x, y (top left corner), ``width'', and ``height'' are obtained from the bounding box of detected faces. 

	Third, the captured pictures become blurry when the rotational speed is high. To prevent it, we utilize information from smartphone sensors to provide feedback ``too fast" when the rotational speed exceeds 20 degrees per second (we found this threshold experimentally). The app will automatically select face images from the camera preview.  The collected images are sent to a Deep Webservice which is responsible for the training/re-training recognition model, versioning trained model, and data. A demo for creating a personal profile is available at link  \footnote{https://youtu.be/Yape97iS-O0 } 
\subsection{Prototype for Smart Door}
\begin{figure}
\includegraphics[width=\textwidth]{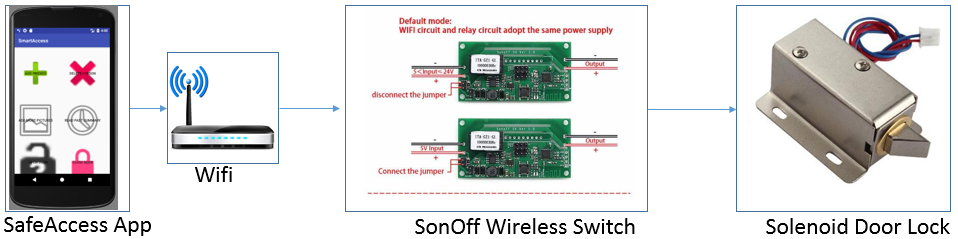}
\caption{ Prototype design of a smart door. From SafeAccess app when users send a command to open the door, then Sonoff switch turn on the lock and its open the door} 
\label{smartdoor}
\end{figure}

One of the vital components of a smart home is a safe and secure door with a smart lock. A smart lock allows the user to enter a house without requiring physical keys. Although there are a lot of commercial smart door systems available, but most of them are very expensive. Hence, we have used Sonoff SV Wifi-enabled switch to control solenoid lock for building a prototype and testing an end-to-end system. It costs only \$12. The underlying architecture of the smart door is shown in figure \ref{smartdoor}. Using SafeAccess app when users send a command to open the door, then Sonoff switch turn on the lock and its open the door. The door is closed automatically after a certain time elapsed or based on user command. This time interval can be set according to users' preferences. Solenoid lock is configured to work in an inverted mode. Therefore, when there is no power in the lock, the door remains closed; that's how we can save power consumption. SonOff runs on a low voltage (5-12V) input power supply. To ensure the highest level of security, Sonoff can be configured to set user-defined wifi SSID (Service Set Identifier) and password. Users will be able to unlock the door from anywhere if Sonoff is connected to home internet wifi. 

\subsection{Feedback Module:} The primary task of the feedback module is sending notifications to users with image descriptions. Although smartphones are easily accessible nowadays, lots of people with a disability do not know how to utilize accessibility features such as TalkBack, Siri, etc. properly. Hence, designing an effective feedback system for people with disabilities is very challenging. Considering the technical adaptability and user preference from participatory design, we have included four types of feedback modes such as MMS, alert with the customized tone, email, and phone call. The feedback mode can be set based on user choice. We developed a communication API using the SMTP (Simple Mail Transfer Protocol) server to send feedback messages to the users via their phone operator. To make a phone call, we are using Twilio \footnote{https://www.twilio.com/} 3rd party service. The average round trip time to receive a messages (generating image description and notification) is 0.15 seconds.    

\section{Model development}
\begin{figure}
\includegraphics[width=\textwidth]{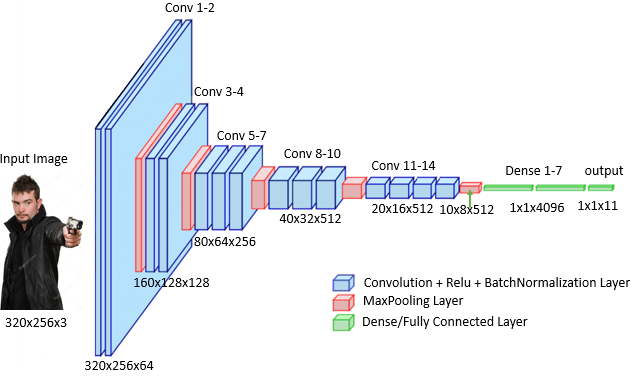}
\caption{SafeNet network architecture. It has 14 convolutional layers, 7 dense layer, 5 maxpooling layers} 
\label{model}
\end{figure}

\begin{figure}
\includegraphics[width=\textwidth]{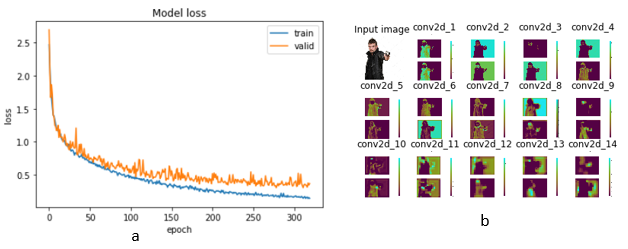}
\caption{a) Training and validation loss curve of SafeNet model. The exponential decay of loss demonstrates the good fitting of the model. b) This figure shows two sample activation maps of SafeNet layer by layer. We can see every neuron learned some distinguishing features and contributed to the final classification} 
\label{model_all}
\end{figure}

We built a Convolutional Neural Network (CNN) based model called “SafeNet” to generate image descriptions. SafeNet has one input layer of dimension 320x256x3, 14 convolution layers, seven dense layers, and 5 MaxPooling layers. The output of each activation is normalized by batch normalization layers. BatchNormalization \cite{ioffe2015batch} helps to prevent covariance shift and model overfitting. Since pooling layers reduce network dimensions very cheaply by discarding lots of spatial information, we used only five MaxPooling layers, which help to reduce the number of parameters and required computational resources. The network architecture and the loss curve are delineated in figure \ref{model}  and \ref{model_all}.a respectively. We came up with this architecture considering some key factors such as the size of the training dataset, overfitting vs. an underfitting problem with network depth and complexity, co-adaptation, feature learning, and predictive ability of individual neurons. The standard network such as VGG16 \cite{simonyan2014very}, ResNet50 \cite{he2016deep}, MobileNet \cite{howard2017mobilenets}, etc. do not perform well with/without transfer learning (see Quantitative Evaluations) for this dataset because simple networks do not learn all distinguishing features and very complex network suffers from the over-fitting problem.

SafeNet has been trained for 320 epochs with a batch size of 24 using Stochastic gradient descent (SGD) optimizer. Finding optimal values for the key parameters (learning rate, momentum) of the SGD optimizer manually/iteratively is exhaustive. Besides, a low learning rate may cause slow convergence, while a high learning rate may miss the global optimum. This issue has been addressed using Baysian Optimization \cite{snoek2012practical}- a method for finding the optimal value of an unknown function. The optimal values obtained from Bayesian optimizer are 0.00101 and 0.8605 respectively from a wide search space of learning rate [0.00001, 0.1] and momentum [0.5, 0.9]. The SafeNet network learned a total of 285,634,121 parameters. There was some criticism in the past regarding the feature learning and their interpretability in the neural network. It is indispensable to have a clear understanding of why any model performs so well, what kind of features that model learned, which part of an image played a significant role in the final classification, whether two independent neurons learned different features and all neurons have the predictive capacity, etc. To understand and explain those scenarios, we have visualized filters, weights, and activity maps of the network layer by layer (see figure \ref{model_all}.b) in input pixels space. We can see from figure \ref{model_all}.b  that the bottom-layers learned edges, blobs, and textures, while upper-level layers learned higher-level abstract. All layers learned distinct features and contributed to the final classification.
\subsection{Data collection}
\begin{figure}
\includegraphics[width=\textwidth]{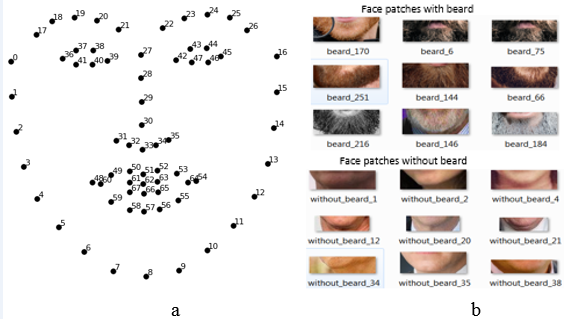}
\caption{a) shows 68 facial landmarks obtained from the face detector. b) shows cropped fac-part from two sample classes (e.g., beard and no-beard)} 
\label{datacol}
\end{figure}

We generated image descriptions by recognizing a set of visible items (gun, knife, scissors, baseball bat, iron bar, eyeglass, mask, cellphone) that incoming persons may carry with and from their facial descriptions. To train SafeNet we collected 8128 image samples from ImageNet \cite{deng2009imagenet}, RGB-D \cite{lai2011large} and web. Then the collected images were sorted out based on the availability of faces with/without a beard, eyeglasses, mustache, and hair. We developed a simple and computationally efficient algorithm (see figure \ref{algs}.a) to crop different parts of a face from collected images by finding facial landmarks \cite{kazemi2014one}  (see figure \ref{datacol}.a) and grouped similar parts (see two sample groups in figure \ref{datacol}.b). Then, three-domain experts examined every single face-part and filtered out parts that are too small (less than 20x20). To make the model affine (rotation, translation, shear, scale) invariant within a certain range, we applied data augmentations so that it can recognize face parts robustly with various orientation and head poses. Data augmentation is a very useful technique to increase the diversity of the data significantly by padding, flipping, rotating, scaling, etc. We  generated a total of 50112 samples for training and validation. 

\section{Results and Discussion} 
The robustness of person detection, recognition, and image description generation model has been evaluated against changes in photometric and geometric properties of images, background clutters, and natural artifacts. We  collected TEST samples from standard datasets, movie clips, and web to examine the computational complexity and generalizations of the models. The collected samples have images of people of various ages, colors, ethnicity, profession, and facial features. Besides, we captured real-time videos using Logitech C270 HD webcam placed in-front of the door to check how the system performs in natural settings. The performance of the model to generate an image description for assessing incoming threats has been tested with movie clips containing crime scenes. To test the end-to-end system, we created a profile using the SafeAccess app that contains 180 images captures from 16 peoples and trained person recognition model. It took 55 and 95 seconds to train the person recognition models built for standalone and integrated mode respectively. The quantitative evaluation for each model is shown in the following sections.

\subsection{Person detection outcomes:}  
\begin{figure}
\includegraphics[width=\textwidth]{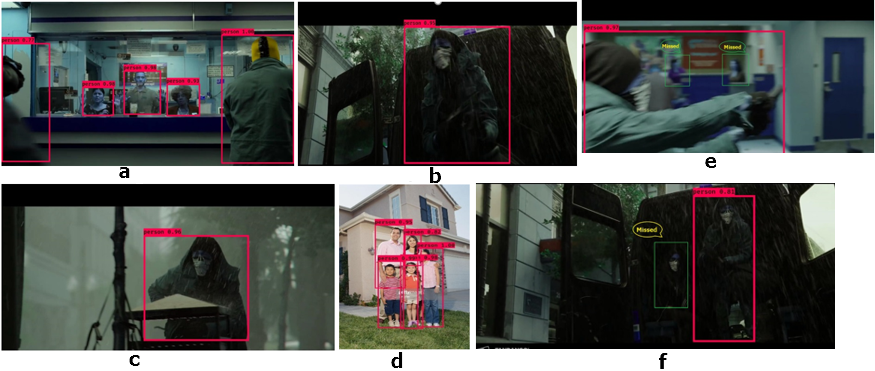}
\caption{Robustness of person detection against: a) complex scene. b) poor lighting condition. c)  natural artifact/rain. d) well-lighted image. and Extreme cases where model failed to detect persons- e) blurry image f) dark image} 
\label{pd_out}
\end{figure}

In figure \ref{pd_out}, some sample outcomes have been shown and we could see from the scene a, b, c (collected from bank robbery movie clip) the person detection model worked robustly against complex scene with background clutters, poor lighting condition and natural artifacts. We categorized the collected video samples into two groups and detection results are presented in Table \ref{t_pd}. Provided the context, the model must be very robust in detecting all incoming persons to the premises regardless of their color, gender, and appearances. Hence, the miss rate/false rejection rate is very important compare to the false acceptance rate and Table \ref{t_pd} shows that the miss rate is very low. However, we found scenarios (see figure \ref{pd_out}.e,f) where the model missed to detect some persons. The reasons for failing are very blurred images (figure \ref{pd_out}.e) and had very poor lighting conditions (figure \ref{pd_out}.f). It is very difficult even for humans to find persons in those scenes.

\begin{table}
\caption{Performance of person detection model with images captured in various conditions}\label{t_pd}
\begin{tabular}{l|l|r|r|}
\hline
 \textbf{Video} &\textbf{Robustness of person } & \textbf{False Rejection} & \textbf{False Acceptance} \\
\textbf{Groups} &\textbf{ detection for images with:} &\textbf{Rate} & \textbf{Rate} \\
\hline

Group 1& Natural artifact, poor lighting,  &	2.44\%	&0.45\%\\
& condition, background clutter, mask & &
\\ \hline

Group 2& person from various color, age, gender&	2.34\%&	0.111\%
\\ \hline

\end{tabular}
\end{table}
 
\begin{figure}
\includegraphics[width=\textwidth]{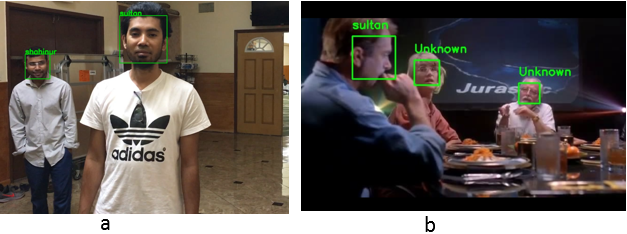}
\caption{Sample result on person identification: a) two known person are identified correctly b) one of the three unknown people identified incorrectly} 
\label{pr_out}
\end{figure}

\begin{table}
\caption{F-measure of personal identification for frontal faces }\label{t_pr}
\begin{tabular}{l|r|}
\hline
 \textbf{Model} &\textbf{F-measure of personal identification} \\ \hline
 Eigenface&0.94 \\ \hline
 FisherFace&0.95 \\ \hline
 LBP/SVM &0.96 \\ \hline
 FaceNet&0.97 \\ \hline

\end{tabular}
\end{table}
\subsection{Person Recognition outcomes}
The quantitative evaluation of the person recognition model showed promising outcomes (see Table \ref{t_pr}) for well-lighted scenes and frontal faces with the limited head pose. However, the model fails to recognize persons with profile face or large head poses (see figure \ref{pr_out}.b). For example, one of the persons in figure \ref{pr_out}.b has a profile face and he was identified incorrectly as a known person. The F-measures for  persons identification is 0.97. A demo is available at link \footnote{ https://youtu.be/J2UizZZKsnY}. To test robustness of model we included person with different facial features such as beard, mustache and wearing eyeglasses, caps etc. in the video.

\subsection{Image Descriptions Generation Outcomes} SafeNet model has been evaluated with four public datasets such as Caltech \cite{Caltech}, UTK \cite{UTKFace}, CelebA \cite{CelebA}, Yale \footnote{http://cvc.cs.yale.edu/cvc/projects/yalefaces/yalefaces.html} and image samples collected from the web and movie clips. SafeNet recognizes items and generates image descriptions with an average F2-measure 0.97. The comparison of performance of SafeNet with VGG16, ResNet50, and MobileNet has been shown in Table 2. A demo on image description generation is available at link \footnote{ https://youtu.be/ICWQwXFf3Kw}

\begin{table}[h]
\caption{Average F-measure of class/item identification to generate image description }\label{t_it}
\begin{tabular}{l|@{}c@{}|@{}r@{}|@{}r @{}|@{}r @{}|r}
\hline
 \textbf{Model}&\textbf{Dataset} &\textbf{Precision}&\textbf{Recall} &\textbf{F-measure} &\textbf{Average F-measure} \\ \hline
SafeNet&
\begin{tabular}{l} Caltech \\ \hline 
UTK \\ \hline
CelebA \\ \hline
Yale \\ 
\end{tabular}
&
\begin{tabular}{r} 0.99\\ \hline
0.99 \\ \hline
0.97 \\ \hline
0.96 \\ 

\end{tabular}
&
\begin{tabular}{r} 0.99\\ \hline
0.97 \\ \hline
0.96 \\ \hline
0.96\\ 

\end{tabular}
&
\begin{tabular}{r}0.99 \\ \hline
0.97 \\ \hline
0.96  \\ \hline
0.96\\ 
\end{tabular}
&
0.97
\\ \hline

ResNet50&
\begin{tabular}{l} Caltech \\ \hline 
UTK \\ \hline
CelebA \\ \hline
Yale \\
\end{tabular}
&
\begin{tabular}{l}
0.87 \\ \hline
0.92 \\ \hline
0.94 \\ \hline
0.86 \\ 
\end{tabular}
&
\begin{tabular}{l}
0.83 \\ \hline
0.91 \\ \hline
0.91 \\ \hline
0.69 \\

\end{tabular}
&
\begin{tabular}{l}
0.84 \\ \hline
0.91 \\ \hline
0.92 \\ \hline
0.72 \\ 
\end{tabular}
&
0.85
\\ \hline

VGG16&
\begin{tabular}{l} Caltech \\ \hline 
UTK \\ \hline
CelebA \\ \hline
Yale \\ 
\end{tabular}
&
\begin{tabular}{l}
0.96 \\ \hline
0.97 \\ \hline
0.98 \\ \hline
0.74 \\ 
\end{tabular}
&
\begin{tabular}{l}
 0.96 \\ \hline
 0.96 \\ \hline
 0.96 \\ \hline
0.96 \\ 

\end{tabular}
&

\begin{tabular}{l}
0.96 \\ \hline
0.95 \\ \hline
0.97 \\ \hline
0.56 \\

\end{tabular}

&
0.86
\\ \hline

MobileNet&
\begin{tabular}{l} Caltech \\ \hline 
UTK \\ \hline
CelebA \\ \hline
Yale \\ 
\end{tabular}
&

\begin{tabular}{l}
0.96 \\ \hline
0.96 \\ \hline
0.97 \\ \hline
0.82 \\ 

\end{tabular}

&
\begin{tabular}{l}
0.85 \\ \hline
0.94 \\ \hline
0.91 \\ \hline
0.67 \\ 

\end{tabular}
&
\begin{tabular}{l}
0.87 \\ \hline
0.94 \\ \hline
0.92 \\ \hline
0.7 \\ 
\end{tabular}
&
0.86
\\ \hline
 
\end{tabular}
\end{table}

\section{Conclusion}
In this paper, we have presented SafeAccess, an end-to-end assistive solution for building safer smart homes and increasing situational awareness. The system monitors homes and classifies incoming persons into friends/family vs unknown/intruders. SafeAccess provides safe and easier access to smart homes and helps people with disabilities to live independently and with dignity. In this work, we focused on gathering user's needs through participatory design, building robust models for person detection, identification and generating image descriptions. The quantitative evaluations and initial user study demonstrated that the SafeAccess enhances the safety and quality of lifestyle of people with disabilities. SafeAccess does not help users to assess threats if an incoming person carries harmful items such as a handgun, knife and keeps it hidden. It is even very challenging for a human observer. However, robbers usually bring heavy and semi-automated guns openly to rob premises. So, if the monitoring cameras are installed outside and inside of a premise SafeAccess helps users to assess incoming threats and to find ongoing crimes. In the alpha version, we included a visually impaired individual in the development cycle of SafeAccess to discover and address issues related to accessibility and usability.  In the beta version, we want to conduct a large-scale user study with a group of people with disabilities at Mid-South Access Center for Technology (Mid-South ACT), Memphis, TN.


\bibliographystyle{splncs04}
\bibliography{Safehome}
\end{document}